\documentclass[twocolumn,prl,superscriptaddress,letter]{revtex4}

\usepackage{graphicx}
\usepackage{amssymb}
\usepackage{amsfonts}

\DeclareGraphicsExtensions{.eps}

\begin{document}

\title{Optical bistability in semiconductor microcavities in the nondegenerate parametric oscillation regime: analogy with the optical parametric oscillator}

\date{\today}

\author{A. Baas, J.-Ph. Karr, M. Romanelli, A. Bramati, and E. Giacobino}
\affiliation{Laboratoire Kastler Brossel, Universit\'{e} Paris 6, Ecole
Normale Sup\'{e}rieure et CNRS,\\
UPMC Case 74, 4 place Jussieu, 75252 Paris Cedex 05, France}

\begin{abstract}
We report the observation of optical bistability in a microcavity pumped at the "magic angle". Experimental evidence is
given in the form of a hysteresis cycle of the nonlinear emission as a function of the pump intensity or the position of
the excitation spot. The results can be well understood with simple theoretical considerations that underline the
fundamental analogy between our system and an optical parametric oscillator.
\end{abstract}

\pacs{71.35.Gg, 71.36.+c, 42.70.Nq, 42.50.-p}

\maketitle

In semiconductor microcavities in the strong coupling regime, the eigenmodes of the system are mixed light-matter fields,
the polaritons. This allows to achieve cavity QED effects involving a light-matter field. Exciting experiments such as
coherent control \cite{Kundermann03a} or squeezing \cite{Karr03a} have been performed using four-wave parametric
scattering of polaritons \cite{Savidis00a,Ciuti00}. Again due to the composite nature of the polaritons of the system,
microcavities can be compared either with Bose-Einstein condensates (BEC), where pure matter waves are involved, or with
optical parametric oscillators (OPO), where the nonlinear medium is excited in the transparency region \cite{Baumberg00}.
The analogy with atomic BECs is most useful in the context of nonresonant pumping experiments, which are motivated by the
realization of a "polariton laser" \cite{Kavokin,Deng03,Andre03}. The analogy with OPOs applies to the case of resonant
pumping at the "magic angle" where two pump polaritons are converted into a pair of "signal" and "idler" polaritons
\cite{whittaker}. The physics of OPOs may be used as a guide for the research of new effects in microcavities, such as the
production of nonclassical states of light \cite{Karr03b}. In this paper, we report the observation of polaritonic
bistability, which is analogous to the optical bistability observed in a detuned triply resonant OPO.

Optical bistability has been predicted \cite{Lugiato88} and observed \cite{Richy95} in a triply resonant OPO (i.e. the
cavity is simultaneously resonant for the pump, signal and idler modes). If the three modes are detuned with respect to
the exact cavity resonance, the condition of oscillation imposes that the detunings of the signal and idler modes
normalized to their HWHM ($\Delta_{s}$ and $\Delta_{i}$ respectively), be equal: $\Delta_{s}=\Delta_{i}=\Delta$. Then the
following condition is required for the existence of bistability :

\begin{equation}
\Delta_{p}\Delta>1
\end{equation}

where $\Delta_{p}$ is the detuning of the pump mode normalized to its HWHM \cite{Fabre90}. Bistability can be evidenced by
observing a hysteresis loop in the variations of the signal intensity vs the pump intensity, or vs the cavity length for a
pump intensity above threshold. Very good agreement has been obtained between experiment and theory \cite{Richy95}.

In semiconductor microcavities pumped resonantly at the "magic" angle, two polaritons of the pumped mode of wave vector
$\mathbf{k_{p}}$ in the plane of the layers scatter coherently to a pair of signal and idler modes of wave vector
$\mathbf{0}$ and $2\mathbf{k_{p}}$ respectively. The analogy with OPOs suggests that bistability should also appear under
certain conditions when the three modes are detuned with respect to the exact resonance condition. While in OPOs the
detunings can be controlled by scanning the cavity length, in microcavities it can be done by moving the excitation spot
over the sample surface, which comes to the same thing due to the slight angle between the cavity mirrors \cite{Baas03}.
We did observe bistability, by scanning either the pump intensity or the position of the excitation spot. The results are
found to be in good agreement with a model similar to those developed for OPOs. This is a completely new type of
bistability in semiconductor microcavities, achieved in the strong coupling regime ; it differs from the one obtained by
excitation of the microcavity at normal incidence (also in the strong coupling regime), which is due to the polaritonic
Kerr effect \cite{Messin01,Karr03a,Baas03}.

\begin{figure}[h]
\centerline{\includegraphics[clip=,width=8.5cm]{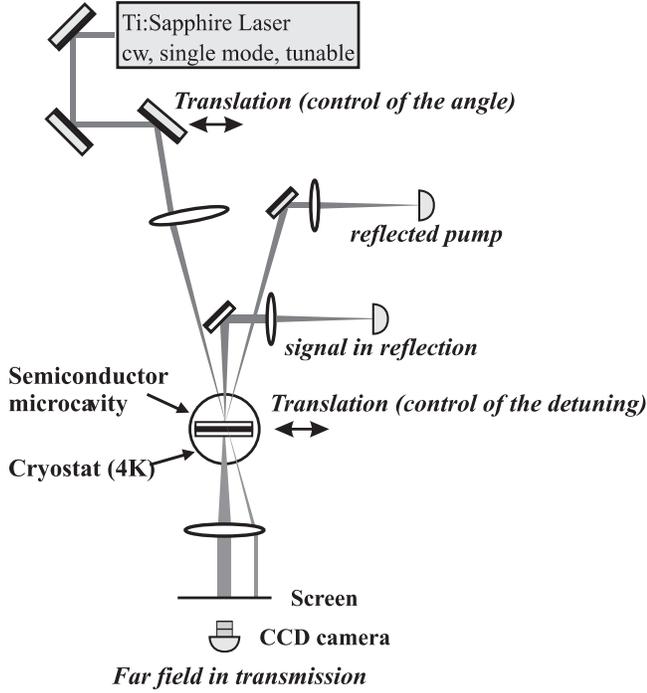}} \caption{Experimental setup. The last mirror before the
focusing lens is mounted on a micrometric translation in the plane of the sample, allowing to vary the angle of incidence
around $12^{\circ}$ without changing the position of the excitation spot.}\label{MontageCPCL}
\end{figure}

The experiments are carried out on a 3$\lambda$/2 microcavity, described in detail in \cite{Stanley98}. The Rabi splitting
is $5.6$ meV and the lower polariton linewidth is of the order of $100$ $\mu$eV. The light source is a single mode,
tunable, intensity stabilized Ti:Sapphire laser. Spatial filtering by a fiber provides a well-defined transverse
distribution of intensity, described by a Gaussian curve with a waist of ($30$ $\pm2$) $\mu$m. The nonlinear emission is
detected by a photodiode in transmission or reflection (see Fig.~\ref{MontageCPCL}). CCD cameras allow to observe the near
field image (image in real space) of the reflected pump beam and of the nonlinear emission. Finally we also used far field
imaging in transmission (image in $\mathbf{k}$-space).

The sample is excited at a cavity-exciton detuning $\delta$ (without any probe beam) at the "magic angle", around
$12^{\circ}$ (depending slightly on the cavity-exciton detuning and the pump intensity), above the parametric threshold
which is of the order of 500 W.cm$^{-2}$. The intensity  of the signal beam emitted around $\mathbf{k}=0$ is optimized, so
that the $\mu$-OPO \cite{Baumberg00} can be considered as perfectly tuned. Then we slightly detune the pump laser by a
quantity $\Delta E$, typically of the order of a few hundreds of $\mu$eV, which results in the disappearance of the
parametric oscillation. In order to obtain it again we have to partly compensate for the pump detuning by moving the
excitation spot on the sample. This comes to changing the cavity length, thus changing simultaneously the frequencies of
the three polariton modes involved ; the direction of the motion must be chosen so as to bring the polariton frequency
back towards the laser frequency.

\begin{figure}[t]
\centerline{\includegraphics[clip=,width=8.5cm]{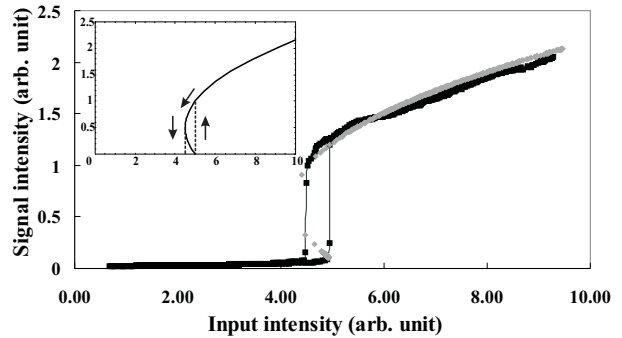}} \caption{Variations of the signal output power (in
transmission) as a function of the pump power for a pump detuning $\Delta E = -0.42$ meV. The gray curve is the result of
a fit using Eq.~(\ref{equationbista}). The inset shows more clearly the unstable branch and the series of intermediate
states that are obtained when varying the input intensity in both directions. The used parameters are $\Delta_{p} = 2.05$
and $\Delta = 1.175$.} \label{HysteresisKlargeP}
\end{figure}

\begin{figure}[t]
\centerline{\includegraphics[clip=,width=8.5cm]{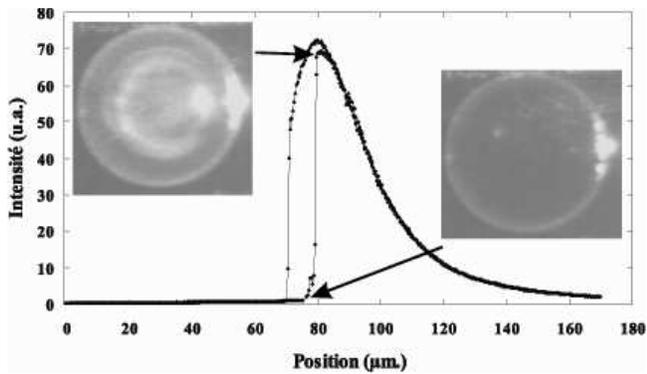}} \caption{Variations of the signal output power (in transmission) as a function of the spot position
in the region of bistability. The pump power is 1150 W.cm $^{2}$ , the pump detuning is $\Delta E = -0.35$ meV. The insets are far field images taken at
the lower and upper point of bistability. The transmitted pump (at 12$^{\circ}$) appears as a bright spot on the right side of the image, in a ring
corresponding to Rayleigh scattering, where speckle patterns are visible. The nonlinear emission appears at an angle of about +6$^{\circ}$ above threshold
(on the same side as the transmitted pump) together with a second Rayleigh scattering ring.} \label{BistaPositionImagesCL}
\end{figure}

For an initial cavity-exciton detuning $\delta = 0$ and a range of laser detunings $\Delta E$ in the window $[-0.14,
-0.42]$ meV, we observed a hysteresis loop in the variations of the signal intensity vs the position of the excitation
spot (for large enough pump intensities). An example is shown in Fig.~\ref{BistaPositionImagesCL}. Then by choosing a
position within the bistable region (corresponding to a new cavity-exciton detuning $\delta^{'}$) one can observe a
hysteresis loop in the variations of the signal intensity vs the pump intensity. An example is shown in
Fig.~\ref{HysteresisKlargeP}. We now show that this bistability regime can be well understood by a simple model of the
parametric process in the microcavity.

The microcavity under resonant excitation at the "magic angle" can be described as a system of three interacting polariton
modes \cite{Ciuti00}, provided it is not too far above the parametric threshold, when multiple scatterings can occur and
more modes have to be taken into account \cite{Savvidis01}. We use a classical model \cite{whittaker} which is sufficient
to describe the main features of the bistability regime.

If the system is perfectly tuned, the signal and idler modes are $\mathbf{k = 0}$ and $\mathbf{k = 2 k_{p}}$ respectively.
But if the system is detuned, the signal and idler wave vectors are not necessarily $\{ \mathbf{0}, \mathbf{ 2 k_{p}} \}$
; the pair of oscillating modes $\{ \mathbf{k}, \mathbf{ 2 k_{p} - k} \}$ is the one with the lowest threshold. This is in
sharp contrast with the case of triply resonant OPOs, where the signal and idler wave vectors are fixed by the cavity. On
the contrary, microcavities have a large angular acceptance and do not impose the signal and idler wave vectors. This
effect is illustrated in the far field images of the nonlinear emission (in transmission) shown in the insets of
Fig.~\ref{BistaPositionImagesCL}. Both images show the transmitted pump and the Rayleigh scattering ring, where speckles
are well resolved \cite{Houdre00}. When crossing one of the bistability turning points, the nonlinear emission suddenly
appears (or disappears) around $6^{\circ}$, well away from the $\mathbf{k = 0}$ direction, together with a ring
corresponding to the Rayleigh scattering as for the pump mode.

In the following, we simply denote the signal and idler modes by the indices $i,s$ ; Their wave vectors $\{ \mathbf{k,2
k_{p} - k} \}$ will be determined later on. The evolution equations for the three polariton modes are

\begin{eqnarray}
\frac{dp_{s}}{dt} &=& -(\gamma _{s} + i \widetilde{E}_{LP}(k_{s}))p_{s} + E_{int}p_{i}^{*}p_{p}^{2}
\label{signal1} \\
\frac{dp_{p}}{dt} &=& -(\gamma _{p} + i \widetilde{E}_{LP}(k_{p}))p_{p}-2 E_{int}p_{p}^{*}p_{s}p_{i} -
C_{p}\sqrt{2\gamma_{a}}A^{in}_{p} \nonumber
\label{pompe1}\\
\frac{dp_{i}}{dt} &=& -(\gamma _{i} + i \widetilde{E}_{LP}(k_{i}))p_{i} + E_{int}p_{s}^{*}p_{p}^{2} \label{idler1}
\end{eqnarray}

where $E_{int}$ is the nonlinear coupling constant between polaritons \cite{Ciuti00} and $\widetilde{E}_{LP}(k_{j})$ the
lower polariton energy at the wave vector of the mode j, taking the renormalization by the pump mode into account. It is
related to the bare polariton energy $E_{LP}(k_{j})$ by $\widetilde{E}_{LP}(k_{j}) = E_{LP}(k_{j}) + 2 \alpha_{j}
|p_{p}|^2$ where $\alpha_{j}$ is the nonlinear coupling constant between the mode $j$ and the pump mode.$\gamma _{j}$ and
$\gamma _{a}$ are the linewidths of the mode j and of the coupling mirror. $A^{in}_{p}$ is the incoming laser field
resonant with the pump mode. We choose the origin of phases so that $A^{in}_{p}$ be real. $C_{p}$ is the Hopfield
coefficient representing the photon fraction of the polariton in the pump mode \cite{note}.

In experiments, the system is first set at resonance for a given cavity-exciton detuning $\delta$, and then detuned by
changing the pump energy (by $\Delta E$) and the cavity-exciton detuning (to a new value $\delta^{'}$). In order to
introduce the corresponding detunings for the three modes, the equation for each mode j is rewritten in the rotating frame

\begin{equation}
\widetilde{p}_{j}(t)=p_{j}(t) \; e^{iE_{j}(\delta^{'})t/\hbar}
\end{equation}

where $E_{j}(\delta^{'})$ is the energy of the mode j for the detuning $\delta^{'}$. We also introduce the normalized
quantities

\begin{eqnarray}
P_{s} &=& \left( \frac{2 E_{int}}{\gamma_{p}}\sqrt{\frac{\gamma _{s}}{\gamma _{i}}} \right) ^{1/2} \widetilde{p}_{s}, \;
P_{i} = \left( \frac{2 E_{int}}{\gamma_{p}}\sqrt{\frac{\gamma _{i}}{\gamma _{s}}} \right) ^{1/2} \widetilde{p}_{i}, \nonumber \\
P_{p} &=& \left( \frac{E_{int}}{\sqrt{\gamma _{s}\gamma _{i}}} \right) ^{1/2} \widetilde{p}_{p}, \nonumber \\
 P^{in}_{p} &=& - \left( \frac{E_{int}}{\sqrt{\gamma _{s}\gamma _{i}}} \right) ^{1/2}
 \frac{C_{p}\sqrt{2\gamma_{a}}A^{in}_{p}}{\gamma_{p}}.
\end{eqnarray}

The stationary regime is given by the following equations :

\begin{eqnarray}
0 &=& -(1+i\Delta_{s})P_{s}+P^{2}_{p}P^{*}_{i} \label{signalStat}\\
0 &=& -(1+i\Delta_{p})P_{p}-P_{s}P_{i}P^{*}_{p} + P^{in}_{p} \label{pompeStat}\\
0 &=& -(1+i\Delta_{i})P_{i}+P^{2}_{p}P^{*}_{s} \label{idlerStat}
\end{eqnarray}

where $\Delta _{j}$ is the detuning of the mode $j$ with respect to the polariton branch normalized to the relaxation
rate: $\Delta _{j}=( \widetilde{E}_{LP} (k_{j}, \delta^{'}) - E_{j} (\delta^{'}) )/\gamma _{j}$. These equations are
similar to those used in the description of a triply resonant OPO. One difference is that two pump photons (instead of
one) are needed for the parametric process, but this does not significantly alter the dynamics of the system. More
importantly, there are additional Kerr-like terms due to renormalization by the pump mode. By combining the equation
(\ref{signalStat}) with the conjugate of the equation (\ref{idlerStat}), one obtains

\begin{eqnarray}
P_{s}[|P_{p}|^{4}-(1+\Delta_{s}\Delta_{i}+i(\Delta_{s}-\Delta_{i}))]=0
\end{eqnarray}

from which we deduce the oscillation condition $\Delta_{s}=\Delta_{i} \label{condoscillation}$, and the pump mode
intensity $|P_{p}|^{2}=\sqrt{1+\Delta^{2}}$, with $\Delta = \Delta_{s} = \Delta_{i}$. Then equations (\ref{signalStat})
and (\ref{idlerStat}) give $|P_{s}|^{2}=|P_{i}|^{2}$. From equation (\ref{pompeStat}) we know that the sum of the phases
of the two final modes is fixed : $\varphi_{s}+\varphi_{i}=0$, where $P_{s}=|P_{s}|e^{i\varphi_{s}}$ and
$P_{i}=|P_{i}|e^{i\varphi_{i}}$. On the contrary, the phase difference can take any value. Making the choice of phase
$\varphi_{s}-\varphi_{i}=0$, one gets $P_{s}=P_{i}$ and the system (\ref{signalStat}-\ref{idlerStat}) reduces to the two
following equations :

\begin{eqnarray}
0 &=& P^{in}_{p}-(1+i\Delta_{p})P_{p}-P^{2}_{s}P^{*}_{p} \label{pompe2}   \\
0 &=& -(1+i\Delta)P_{s}+P^{2}_{p}P^{*}_{s} \label{signal2}
\end{eqnarray}

The equations for the stationary state of the four wave mixing are found to be formally the same as in the case of
frequency degenerate four-wave mixing ($P_{s}=P_{i}$). The solutions are given by

\begin{equation}
|\overline{P}^{in}_{p}|^{2}=(1+\Delta^{2})|\overline{P}_{s}|^{4}+2(1-\Delta_{p}\Delta)|\overline{P}_{s}|^{2}+
1+\Delta^{2}_{p} \label{equationbista}
\end{equation}

with $\overline{P}^{in}_{p} = P^{in}_{p} (1+\Delta^{2})^{-1/4}$ and $\overline{P}_{s} = P_{s} (1+\Delta^{2})^{-1/4}$. This
equation is similar to the equation giving the stationary solutions in a triply resonant OPO \cite{Lugiato88}. In our case
however, the detunings $\Delta$ and $\Delta_{p}$ depend on the polariton population in the pump mode due to
renormalization effects. Since the pump mode population is clamped at a fixed value above the parametric threshold, this
only results in a constant shift of $\Delta$ and $\delta_{p}$ with respect to the equations of the OPO.

As a consequence we obtain the same conditions for the existence of a bistability regime. If $\Delta_{p}\Delta<1$, there
is at most one stable stationary state for each value of the input intensity. For $\Delta_{p}\Delta>1$, bistability
appears in the interval

\begin{eqnarray}
\frac{(\Delta_{p}+\Delta)^2}{1+\Delta^{2}}<|\overline{P}^{in}_{p}|^{2}<1+\Delta^{2}_{p} \label{seuilbista}
\end{eqnarray}

Fig.~\ref{HysteresisKlargeP} shows a fit of the experimental curve with the theoretical curve given by
Eq.~(\ref{equationbista}). Very good agreement is obtained. The theoretical curve shows an additional negative-slope
branch in the bistability region, which can be proved to be unstable. Thus the signal turns on and off at the two turning
points, as schematized in the inset of Fig.\ref{HysteresisKlargeP}.

The model also allows to calculate the wave vectors of the signal and idler modes $\{ \mathbf{k,2 k_{p} - k} \}$. To do
this, we minimize the oscillation threshold $|P_{p}^{in}|^{2}=\sqrt{1+\Delta^{2}} \; (1+\Delta_{p}^{2}$ as a function of
$\mathbf{k}$. One finds out that when the system is detuned, the signal-idler pair having the lowest threshold is no
longer $\{ \mathbf{0}, \mathbf{ 2 k_{p}} \}$. For the parameters of Fig.\ref{BistaPositionImagesCL}, the angle of
propagation of the signal beam is found to shift by about +4$^{\circ}$. This is in good qualitative agreement with the
images in Fig.~\ref{BistaPositionImagesCL}, where the angle of emission of the signal beam is about +6$^{\circ}$.
Interestingly, this behavior differs from the one observed in \cite{butte,gippius} where the signal occurs at
$\mathbf{k=0}$ whatever the excitation conditions.

Finally let us comment on the parameters of observation of bistability. We did not observe bistability for a laser
detuning -0.14 $<\Delta E <$ 0 meV. This can be explained with the bistability condition $\Delta_{p}\Delta>1$ : the pump
has to be sufficiently detuned. We did not observe bistability for large detunings such as $\Delta E <$ -0.42 meV, which
is probably due to the fact that the bistability threshold actually increases more with $\Delta E$ than predicted by the
model. In order to get a better agreement with experiments, the model has to be refined by including the precise
dependence of the polariton linewidths as a function of the cavity-exciton detuning $\delta$ and the wave vector.

Thus the main features of this bistability regime are reproduced by a model adapted from the theory of bistability in
OPOs, that also allows to understand previous experimental results obtained by R. Houdr\'{e} \textit{et al.} on the same
sample \cite{Houdre00}. They observed a very steep thresholdlike behavior of the emission intensity around $\mathbf{k}$=0
vs the pump intensity, strongly deviating from the standard shape of the parametric oscillation threshold curves. By
taking the parameters in which this curve was obtained, we have checked that the system does present bistability because
it is detuned with respect to the optimal conditions of parametric oscillation.

A further point requiring investigation is the transverse shape of the nonlinear emission. While the images in the inset of
Fig.~\ref{BistaPositionImagesCL} are well understood in the frame of our model, by varying the experimental conditions one observes complex patterns
changing with the cavity-exciton detuning and the excitation intensity. Very similar results were obtained in Refs.~\cite{Houdre00,Houdre00b}. In
Ref.\cite{Baas03} we have developed a simple model of transverse effects which gives satisfactory results in the degenerate configuration of the
parametric process, showing that the transverse patterns can be attributed to a modification of the shape of the active region, due to the combined effect
of exciton renormalization, the angle between the cavity mirrors and the gaussian distribution of the excitation spot. A similar study in the "magic
angle" geometry would be an important step forward since transverse effects should certainly be taken into account in the evaluation of the polariton
density and the polariton occupation number.

As a conclusion we have reported the observation of bistability, which is a new feature of the nonlinear dynamics of
polaritons in the parametric oscillation regime. Experimental results are in good qualitative agreement with a model
adapted from the theory of bistability in OPOs. Further studies should be devoted to transverse aspects. Experiments under
pulsed excitation would also provide interesting information on the dynamics of polariton bistability, which should be
very different from the dynamics of OPOs, because real (and not virtual) material excitations are involved in the
parametric process \cite{Kundermann03b}. Finally this work confirms the robustness of the analogy between microcavities
and OPOs, at the base of the prediction of quantum correlated light-matter waves \cite{Schwendimann,Karr03b}.

We would like to acknowledge fruitful discussions with C. Fabre, L. Longchambon, S. Kundermann and N. Treps. We thank R.
Houdr\'{e} for providing us with the microcavity sample.


\begin{thebibliography}{99}
\bibitem{note}All the parameters -polariton energies and linewidths, interaction constant and Hopfield coefficient- depend
on the cavity-exciton detuning, but this dependence is not noted for the sake of clarity.
\end{thebibliography}
\end{document}